\documentclass[twocolumn,twocolappendix]{aastex631}
\graphicspath{{./}{Figures/}}
\usepackage{amsmath}

\usepackage{rotating}
\usepackage{afterpage}
\usepackage{float}

\begin{document}

\title{Ergodicity of FIRE: star formation variations within and between simulated galaxies}

\author{Fraser M. Smith}
\affiliation{Department of Astronomy and Physics and Institute for Computational Astrophysics \\ 
Saint Mary's University \\ 
Halifax, NS B3H 3C3, Canada}

\author{Robert J. Thacker}
\affiliation{Department of Astronomy and Physics and Institute for Computational Astrophysics \\ 
Saint Mary's University \\ 
Halifax, NS B3H 3C3, Canada}



\begin{abstract}

We investigate the ergodicity of star formation in simulated galaxies from the FIRE-2 (Feedback In Realistic Environments) project. We restrict ergodicity considerations to being related to deviations from the star-forming main sequence (SFMS), and in turn whether ensemble averages across populations match time-averaged star formation histories (SFHs) based on simulated observable properties. We find that in these high-resolution simulations the deviations of individual galaxies from the SFMS tend to approach ergodic behavior over time, regardless of the SFMS definition adopted and the star formation estimator used. This trend persists when galaxy morphology, as traced by S\'ersic index, is considered despite the spheroid-dominated morphologies showing a smaller range of SFMS deviations than the disk-dominated morphologies. Unsurprisingly, we find more rapid convergence to ergodic behavior for star formation estimators based on shorter time scales ($10^{7}$ years) as opposed to longer ($10^{9}$ years). We caution that these findings should be considered in the context of the current sample and that further studies, particularly of high redshift evolution and the impact of active galactic nuclei should be investigated.

\end{abstract}


\section{Introduction}

In the lambda cold dark matter ($\Lambda$CDM) paradigm, dark matter halos form through the gravitational collapse of overdensities. Through hierarchical merging, these halos can grow to the sizes observed at present. How galaxies continue to evolve is dependent on many factors, for example, the environments in which the galaxies are situated \citep[e.g.][]{mundy2017,laigle2018,singh2024}. Theoretically, if evolution could be predicted based on sample variance, galaxy evolution would be simpler. While general evolutionary trends can be broadly inferred through analyses of samples of galaxies \citep[e.g.][]{kartaltepe2023,lee2024} and spectral energy distribution (SED) fitting \citep{leja2019}, these approaches typically impose uncertainties and biases including underestimated SFR uncertainties and overestimated SFRs in starburst events that further complicate interpretations \citep{haskell2024}. In order to directly infer galactic properties, specifically star formation properties such as their star formation histories (SFHs), from a single observed point in a galaxy's evolution would require that SFHs are ergodic. That is, ensemble averages of SFHs at any given time are representative of/equivalent to individual average SFHs. If ergodicity does not hold, ergodicity breaking occurs which can be classified as weak or strong. The breaking is said to be weak if the system fails to explore the parameter space over a finite time, while strong breaking refers to non-ergodicity due to regions of the parameter space being physically exclusive to each other \citep{spiechow2016}. Given the complex, multi-scale processes involved in galaxy evolution, one does not expect that SFHs would be ergodic, especially when considering the impact of mergers. However, if the deviations from ergodicity that galaxies exhibit can be constrained and quantified, profound insights into the processes driving this non-ergodicity can be made which in turn allows us to infer how galaxies deviate from ergodicity. If we reduce the star formation properties under consideration down to deviations from the main sequence, then the possibility of ergodic behaviours increases. In essence, considering deviations reduce the dimensionality by removing the trend. If galaxies do evolve around the star-forming main sequence (SFMS) in a seemingly stochastic manner, each galaxy's star formation behaviour is expected to match the sample average SFMS. This is particularly true at later times and higher halo masses, where SFHs tend to be less variable and hence galaxies should converge to SFMS deviations close to 0 dex.

Despite complex evolutionary histories and processes that influence galactic evolution, typically galaxies are labeled as either 'star-forming' or 'quiescent'. This classification is based on their relative position to the observed tight relation between SFR and stellar mass, known as the star-forming main sequence (SFMS) of galaxies \citep{noeske2007,schreiber2015,barro2019}. Scatter around this relation comes from fluctuations on both short-term and long-term timescales, with assembly bias accounting for a significant portion of this environmentally-correlated scatter \citep{matthee2019}. In this study, we focus on star-forming galaxies since we are interested in the relative effects of physical processes on star formation, and to measure the ergodicity of deviations from the mean/median of galaxy samples.

Accretion is arguably the primary driver of galactic evolution. This accretion can come from the surrounding circumgalactic medium (CGM) or via galaxy-galaxy interactions. Mergers can induce morphological changes in galaxies \citep[e.g.][]{hopkins2010,clauwens2018}, in addition to altering the galactic specific star formation history (sSFH), depending on the nature of the merger \citep[e.g.][]{athan2016,martin2018,graham2024}. While major mergers have a more pronounced impact on galaxy evolution such as the disruption of stellar disks and contributing significantly to ex-situ stellar mass growth \citep[e.g.][]{rodriguez2016}, minor mergers can also significantly impact galactic evolution \citep[e.g.][]{kaviraj2014}. Minor mergers can produce the strongest disk heating in galaxies, given sufficient (log$_{10}$(M$_{*}$/M$_{\odot}$) $\gtrsim$ 10) mass \citep{kazantzidis2008,grand2016}. Morphology is important to consider when studying star formation trends, and so a distinction must be made between considering disc-dominated and spheroidal systems. Thus, deviations form the SFMS are likely to be morphology-dependent.

Various internal factors also regulate the star formation within a galaxy, and can potentially temporarily remove galaxies from the SFMS \citep[e.g.][]{tacchella2016,leroy2024}. This includes thermal feedback (e.g. Type-Ia and Type-II supernovae, photoionization and background radiation), kinetic feedback in the form of stellar winds, and the less understood feedback from active galactic nuclei (AGN) that can produce thermal and kinetic feedback \citep{weinberger2018,torrey2020,poit2023}. Due to the differing characteristic timescales involved in these processes, it can be difficult to disentangle the precise contributions of each process to the overall star formation present at any given time \citep{iyer2020,iyer2024}. However, the interplay between processes could introduce stochasticity which would increase ergodic behaviour.

Recent data from \textit{JWST} has provided invaluable insight into the complexities of galaxies formed in the early Universe. This includes constraints on previously estimated properties, such as the UV luminosity function \citep{willott2024}, and the presence of bright compact objects at high-redshifts \citep{finkelstein2023}. Surprising evidence has emerged for recently quenched galaxies early in the Universe \citep[e.g.][]{endsley2024} based on large Balmer breaks present at these high redshifts. Certain mass estimates have even brought fundamental theories of galactic mass growth to question \citep[e.g.][]{labbe2023}, although factors such as cosmic variance and the effect of emission on mass-to-light ratios remain important to consider \citep{desprez2024}. Given the recent evidence of higher SFRs and the associated feedback one would link to this, it is interesting to investigate possible ergodicity of high redshift star formation properties. Indeed, ergodicity may be expected at early cosmic times due to the bursty nature of SFHs.

Many galaxy properties are correlated with redshift. Perhaps most notably, the overall sizes of galaxies are known to change with redshift, with galaxies typically being smaller at higher redshifts \citep{baggen2023}. While mergers contribute to the overall growth necessary to match present galaxy sizes, other processes such as adiabatic expansion are required \citep{damjanov2009}. There is also evidence to suggest that sizes of disc galaxies may change significantly at early cosmic times \citep{zolotov2015}. Various well-known relations change with redshift as well, including the SFMS \citep{noeske2007,speagle2014}. The overall SFR density is known to evolve over cosmic time, peaking around z$\sim$2 \citep{madau2014}. Higher merger rates contribute to high-redshift galaxies typically exhibiting more short-term variations in SFHs than galaxies at low-redshifts. With more `bursty' and variable SFHs, we expect galaxies to exhibit more ergodic properties because galaxies are more likely to explore the SFR-M$_{*}$ parameter space.

Due to the long evolutionary time scales involved, investigating ergodicity of any galaxy property requires the use of simulations. By the nature of ergodicity, accurate histories of properties over time are necessary to test this assumption. As more observational data of high-redshift galaxies has improved \citep[e.g.][]{harikane2023}, so too has our ability to reproduce observations through cosmological simulations \citep{vogelsberger2020}. Possible solutions for outstanding observational questions, such as the too-big-to-fail and missing-satellites problems, have been found through simulations \citep[e.g.][]{wetzel2016,garrison2019b}. By utilizing simulations with differing (subgrid) physics implementations, we can better understand the processes involved in driving evolution \citep[e.g.][]{crain2015}. Different simulation codes use various hydrodynamical solvers, and although all methods have differing strengths, certain algorithmic limitations can introduce artificial noise into results, depending on the desired analysis and scales involved \citep[e.g.][]{vandenBosch2018}. For example, while there have been many improvements to the traditional smoothed particle hydrodynamics \citep[SPH,][]{gingold1977} including pressure-based implementations \citep{lind2012}, SPH is inaccurate at subsonic scales which can produce large errors \citep{bauer2012,hopkins2015}. All of these issues may impact the star formation and thus overall conclusions regarding galaxy star formation properties and any apparent ergodicity. We refer to systems that give metric results suggestive of ergodicity, but do not necessarily exhibit true ergodic properties to have ``apparent ergodicity". This is an inherent limitation of the metric which we discuss explicitly in Section~\ref{tm_limits}.

This work builds off the analysis in \cite{smith2024} that was motivated by \cite{wang2020b}. By utilizing state-of-the-art simulations with more realistic physics and more evolutionary factors considered we extend the analysis significantly over prior work. However, we note that the sample used is driven by available simulations and is not constructed by mimicking observational selection. The sample used in this paper is not statistically random, but provides valuable insight for galaxies in an ergodic context. To test discrepancies with true ergodicity, we compare SFHs of individual galaxies to multiple definitions of the SFMS over cosmic time. In Section~\ref{Methods}, the simulations analyzed are described. In Section~\ref{Results}, we summarize our findings before our concluding remarks in Section~\ref{Conclusions}.

\section{Methods}
\label{Methods}
\subsection{FIRE-2 Simulations}
We analyze simulations from the core suite of the publicly-available \textsc{FIRE}-2 simulations \citep{wetzel2023,wetzel2025}. The FIRE-2 cosmological zoom-in simulations of galaxy formation are part of the Feedback In Realistic Environments (FIRE) project, generated using the \textsc{gizmo} code \citep{hopkins2015} and the FIRE-2 physics model \citep{hopkins2018}. This collection of 20 simulations includes 3 Local Group-like galaxy pairs, 8 isolated Milky Way-mass galaxies, 5 intermediate-mass galaxies, and 4 low-mass galaxies. The halos analysed in this work are summarized in Table~\ref{halotab} with references therein. We provide a brief overview of the key features and implementation of physical processes here.

FIRE-2 simulations are run with the meshless finite mass (MFM) hydrodynamical solver \citep{hopkins2015}. Star formation is implemented via the sink-particle approach in gas that is self-gravitating, self-shielding, Jeans unstable, and dense ($n_{H} > 1000$ cm$^{-3}$). Feedback from Type-Ia and Type-II supernovae, continuous mass-loss from AGB and OB stars, photoionization, photoelectric heating, and radiation pressure are implemented. The typical baryonic mass resolution of 7100 M$_{\odot}$ $h^{-1}$ and baryonic force softening scales that can reach below 100 pc allow for the formation and dissipation of giant molecular clouds to be modeled and resolved.

\begin{table*}
\centering
\begin{tabular}{|c c c c c|} 
\hline

Simulation Name & log$_{10}$(M$_{*}$ M$_{\odot}^{-1}$ ) & R$_{200c}$ [kpc] & m$_{baryonic}$ [M$_{\odot}$ $h^{-1}$] & Reference \\

\hline

m11b & 7.606 & 68.903 & 2.6 x 10$^{3}$ & \cite{chan2018} \\
m11d & 9.706 & 132.725 & 7.1 x 10$^{3}$ & \cite{elbadry2018} \\
m11e & 9.221 & 105.405 & 7.1 x 10$^{3}$ & \cite{elbadry2018} \\
m11h & 9.607 & 116.319 & 7.1 x 10$^{3}$ & \cite{elbadry2018} \\
m11i & 9.015 & 83.597 & 7.1 x 10$^{3}$ & \cite{elbadry2018} \\
m11q & 8.870 & 103.895 & 7.1 x 10$^{3}$ & \cite{hopkins2018} \\
m12b & 11.000 & 215.094 & 7.1 x 10$^{3}$ & \cite{garrison2019a} \\
m12c & 10.835 & 213.111 & 7.1 x 10$^{3}$ & \cite{garrison2019a} \\
m12f & 10.986 & 225.762 & 7.1 x 10$^{3}$ & \cite{garrison2017b} \\
m12i & 10.862 & 201.702 & 7.1 x 10$^{3}$ & \cite{wetzel2016} \\
m12m & 11.123 & 222.586 & 7.1 x 10$^{3}$ & \cite{hopkins2018} \\
m12r & 10.388 & 203.641  & 7.1 x 10$^{3}$ & \cite{samuel2020} \\
m12w & 10.818 & 198.049 & 7.1 x 10$^{3}$ & \cite{samuel2020} \\
m12z & 10.384 & 177.127 & 4.2 x 10$^{3}$ & \cite{garrison2019a} \\
Romeo & 10.902 & 219.355 & 3.5 x 10$^{3}$ & \cite{garrison2017b} \\
Juliet & 10.666 & 200.739 & 3.5 x 10$^{3}$ & \cite{garrison2017b} \\
Romulus & 11.074 & 241.397 & 4.0 x 10$^{3}$ & \cite{garrison2019b} \\
Remus & 10.737 & 204.579 & 4.0 x 10$^{3}$ & \cite{garrison2019b} \\
Thelma & 10.922 & 214.654 & 4.0 x 10$^{3}$ & \cite{garrison2017b} \\
Louise & 10.500 & 195.045 & 4.0 x 10$^{3}$ & \cite{garrison2017b} \\
\hline
\end{tabular}
\caption{Overview of FIRE-2 simulations studied. The stellar mass at z = 0 contained within the virial radius, the virial radius at z = 0, and the smallest particle resolution are reported. Note that the stellar masses and virial radii do not exactly match previous studies \citep[e.g.][]{hopkins2018} since we adopt different definitions.}
\label{halotab}
\end{table*}

\afterpage{\clearpage}

\subsection{Quantifying Stellar Mass and Star Formation in Simulated Galaxies}

We define the total SFR of a galaxy at a given time to be the sum of stellar mass formed within the virial radius (R$_{200c}$, the radius at which the enclosed density is 200$\rho_{crit}$) of its host halo. To calculate the SFR, stellar mass formed between snapshots is averaged over the time since the last output. Similarly, stellar mass is defined as the total mass of stellar particles contained within the virial radius. Our definition follows the theoretical motivation of considering all particles gravitationally bound within the galaxy  \citep[e.g.][]{pillepich2018}. These definitions are common when considering simulated galaxies \citep[e.g.][]{guedes2011,sorini2022}.

We note that if we were to only consider the visible stellar mass within the disk, as opposed to all stellar material within the virial radius, our results would not be impacted significantly. That is, although our stellar mass definition that aligns with standard practice in galaxy simulations does not follow an observationally-motivated stellar mass definition, our conclusions remain the same. For example, the average stellar mass difference between mass contained within the virial radius and 10 $R_{e}$ at z = 0 is 0.04 dex. The vast majority of the total SFR and stellar mass at any given time is dominated by the innermost regions of the galaxy. Thus, even using the stellar half-mass radius as our cutoff, we get similar SFRs and the logarithmic nature of the SFMS means that even a factor of 2 difference in stellar mass does not change the magnitude of SFMS deviations in any appreciable way. We note that further restricting our radial extent to 3 $R_{e}$ as is commonly done in observational studies would yield similar results: the main baryonic components between 3 $R_{e}$ and 10 $R_{e}$ are the stellar halo and possibly an extended part of a disk, where an insignificant fraction of star formation is present. The median difference in star formation enclosed between these two radii is 6.5 per cent across all galaxies in our sample, both extended and compact systems.

\subsection{Archaeological Star Formation Histories}

The SFHs of galaxies can also be considered as histograms of stellar formation times in `archaeological' SFHs. Instead of considering all stellar mass that forms between simulation snapshots, stars formed throughout the merger tree are considered, including both ex-situ and in-situ star formation. This provides an upper limit to the stellar mass assembly histories of our sample.

Consistent with previous works \citep[e.g.][]{gurvich2023}, we bin stellar formation times by 1 Myr intervals. We note that the produced SFHs are inherently smooth due to the high sampling rate. Since the archaeological SFRs are sampled in 1 Myr bins compared to averaging SFRs between snapshots, we consider the cumulative stellar mass of galaxies to compare the respective SFHs. Overall, stellar mass is accumulated at a similar rate in both methods, with slightly more mass (on average 0.12 dex) in the archaeological approach since it includes stars formed outside the parent halo. 

We find that our results are similar whether using archaeological SFHs or constructed using each snapshot. That is, the conclusions regarding the apparent ergodic convergence in Section~\ref{tm_metric_sect} for our full sample are the same regardless of using archaeological or calculated SFHs. However, it is important to make a distinction between archaeological SFHs and constructed SFHs: while we find our results are comparatively unaffected, we do encourage other researchers to examine this issue within their own sample.

\subsection{Morphological Considerations}

In order to separate disc and spheroidal galaxies, we emulate traditional observational approaches to characterizing morphologies. Since simulations use mass instead of luminosity for stellar material, we can not apply a S\'ersic cut directly to our data without making assumptions about mass-to-light ratios. While fitting light profiles on mock images may produce well-motivated results, this approach is computationally expensive to perform over many snapshots. Thus, since we are only interested in the slope of the S\'ersic profile, we fit a S\'ersic profile to the surface mass density along the principal axes of the host halos similar to \cite{croft2009}. Previous work has considered the relation between surface brightness and surface mass density \citep[e.g.][]{bovy2013}, and have found that at least for Milky Way-like systems, this substitution is well-motivated considering the near constant mass-to-light ratios seen in the NIR to IR for disk systems \citep[e.g.][]{gavazzi1993,mcgaugh2014}.

S\'ersic profiles are fit using 40 linearly spaced radial bins from 0.25R$_{e}$ to 10R$_{e}$, where we define the effective radius R$_{e}$ of a galaxy to the stellar half-mass radius (R$_{1/2}$). To distinguish disc-dominated versus spheroid-dominated systems, we set a threshold S\'ersic index of $n = 2.5$ consist with previous studies \citep[e.g.][]{barden2005}.

\begin{figure*}
\centering
\includegraphics[scale=0.85]{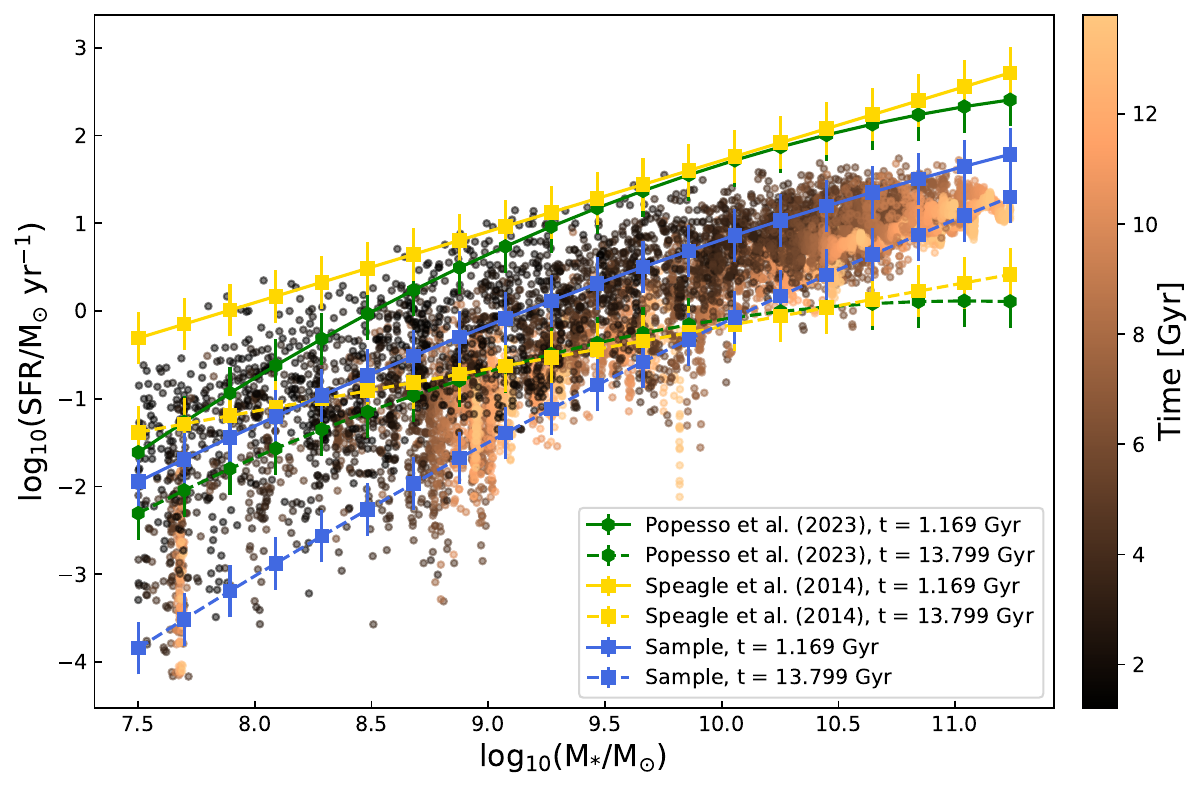}
\caption{SFMS relation for all galaxies analyzed. Simulation results are plotted for all z = 4.0 to z = 0.0 values. The expected SFMSs from \cite{popesso2023} and \cite{speagle2014} are displayed in green and yellow, respectively. The SFMS constructed from the FIRE-2 data is shown in blue. Solid lines indicate the SFMS calculated at z = 4.0 and dashed lines show the expected SFMSs at z = 0.0. Error bars show the accepted scatter of 0.3 dex. 
\label{sfms_all}}
\end{figure*}

\subsection{Defining the SFMS}

How the SFMS is defined dictates how deviations are calculated. We consider three definitions for the SFMS: two observationally-motivated, redshift-dependent SFMSs from \cite{speagle2014} and \cite{popesso2023} as well as a SFMS calculated at each timestep based on the simulated data. By fitting our own SFMS, we create a fit that is more consistent with the simulation sample.

The best-fit SFMS presented in \cite{speagle2014} has the form,
\begin{equation}
\begin{split}
\rm{log}(\rm{SFR}/\rm{M_{\odot}} \rm{yr}^{-1}) =
(\alpha_{c} + \alpha_{t} \times t) \rm{log}(\rm{M}_{*}/\rm{M}_{\odot})\\ 
- (\beta_{c} + \beta_{t} \times t),
\end{split}
\end{equation}
\noindent
where t is the age of the Universe in Gyr, $\alpha_{t} = -0.026 \pm 0.003$, $\alpha_{c} = 0.84 \pm 0.02$,$\beta_{t} = 0.11 \pm 0.03$, and $\beta_{c} = -6.51 \pm 0.24$.

\cite{popesso2023} uses data past 2014 spanning a wide redshift (0 $<$ z $<$ 6) and stellar mass range ($10^{8.5}$ - $10^{11.5}$):

\begin{equation}
\label{pope_eq}
\begin{split}
\rm{log}(\rm{SFR}/\rm{M_{\odot}}  \rm{yr}^{-1}) = & 
(b_{2})\rm{log}^{2}(\rm{M}_{*}/\rm{M}_{\odot}) \\
& + (b_{1} + a_{1} t) \rm{log}(\rm{M}_{*}/\rm{M}_{\odot}) \\
& + (b_{0} + a_{0} t),
\end{split}
\end{equation}

\noindent
where t is the age of the Universe in Gyr, $b_{2} = -0.1925 \pm 0.0011$, $b_{1} = 4.722 \pm 0.012$, $a_{1} = - 0.034 \pm 0.002$, $b_{0} = -26.134 \pm 0.015$, and $a_{0} = 0.20 \pm 0.02$.

To construct a SFMS from the simulated data, we follow the approach in \cite{popesso2023}. Snapshots of all simulations at each time are used to fit Equation~\ref{pope_eq}, where $a_{0}$, $b_{0}$, $a_{1}$, $b_{1}$, and $b_{2}$ are free parameters. In following \cite{popesso2023}, we do not apply mass or redshift binning.

\section{Results}
\label{Results}

\subsection{SFMS Evolution}

Figure~\ref{sfms_all} shows the relation between the SFR and stellar mass for the analysed galaxies at each snapshot. The analyzed galaxies broadly follow observed SFMS trends throughout cosmic time. At lower-masses there is a larger scatter about the SFMS compared to the small scatter seen at higher-masses. This is a reflection of low-mass galaxies exhibiting bursty SFHs as opposed to the smooth, low-variation SFHs seen at high-masses. In addition, this is caused by the sampling effect of selecting high-mass disks which produce extended periods of passive star formation.

\begin{figure}

\centering
\includegraphics[scale=0.33]{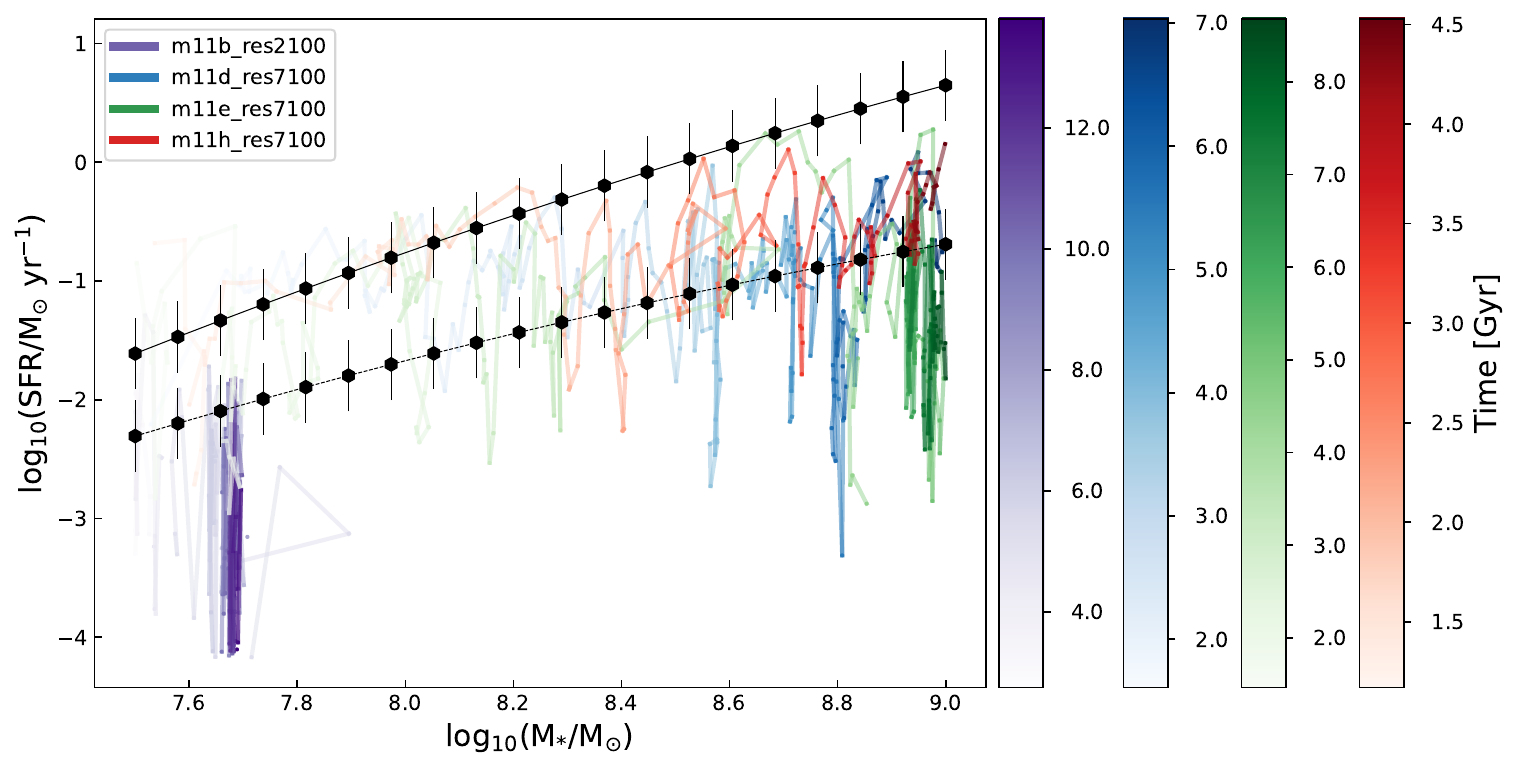}
\caption{Evolution of galaxies in log(SFR/M$_{\odot}$ yr$^{-1}$)-log(M$_{*}$/M$_{\odot}$) over time in the low-mass regime. Black lines correspond to the \cite{popesso2023} SFMS fit. Error bars show the accepted scatter of 0.3 dex.
\label{sfms_track_lm1}}
\end{figure}

\begin{figure}
\centering
\includegraphics[scale=0.33]{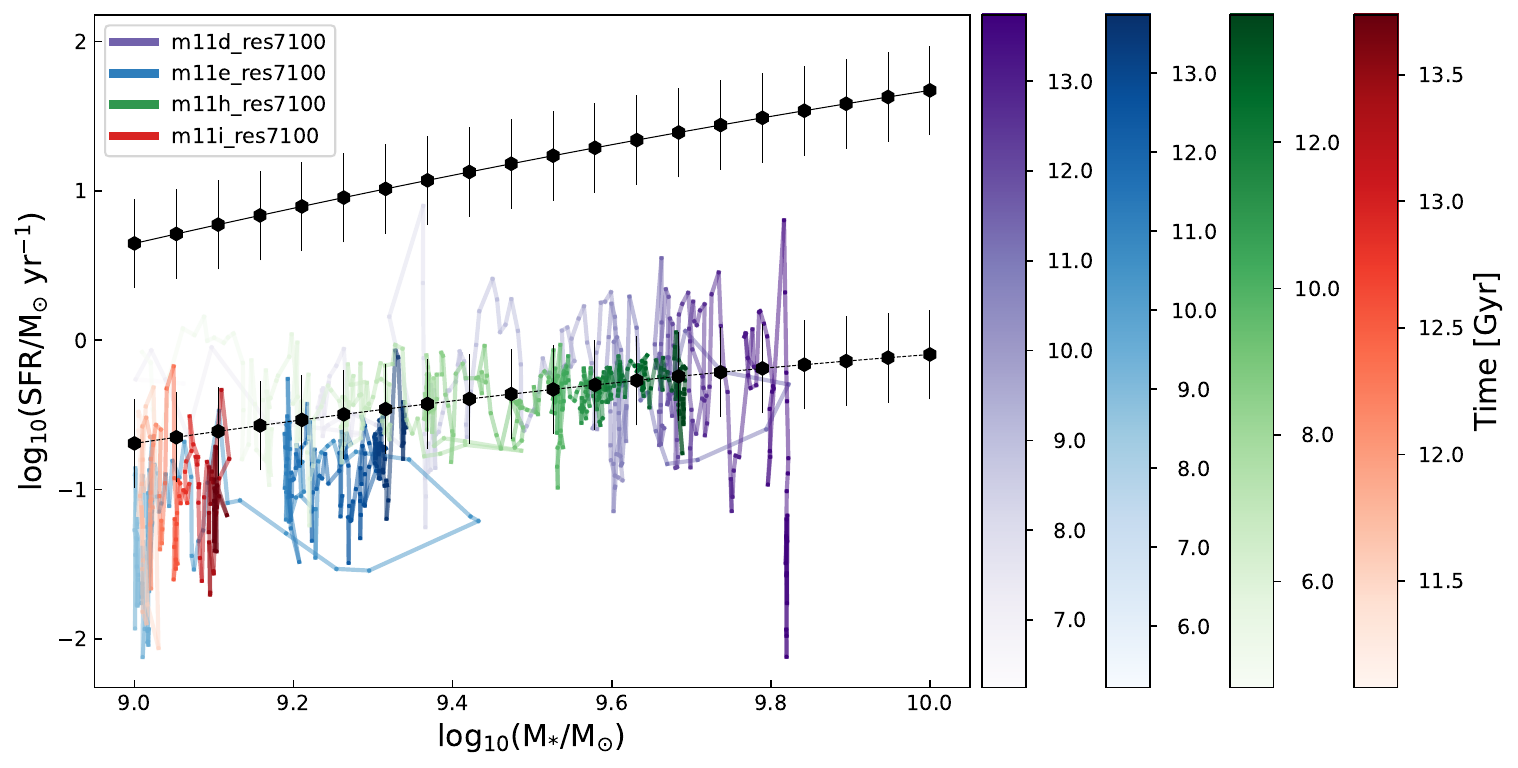}
\caption{Evolution of galaxies in log(SFR/M$_{\odot}$ yr$^{-1}$)-log(M$_{*}$/M$_{\odot}$) over time in the intermediate-mass regime. Black lines correspond to the \cite{popesso2023} SFMS fit. Error bars show the accepted scatter of 0.3 dex.
\label{sfms_track_im1}}
\end{figure}

\begin{figure}[t!]
\centering
\includegraphics[scale=0.33]{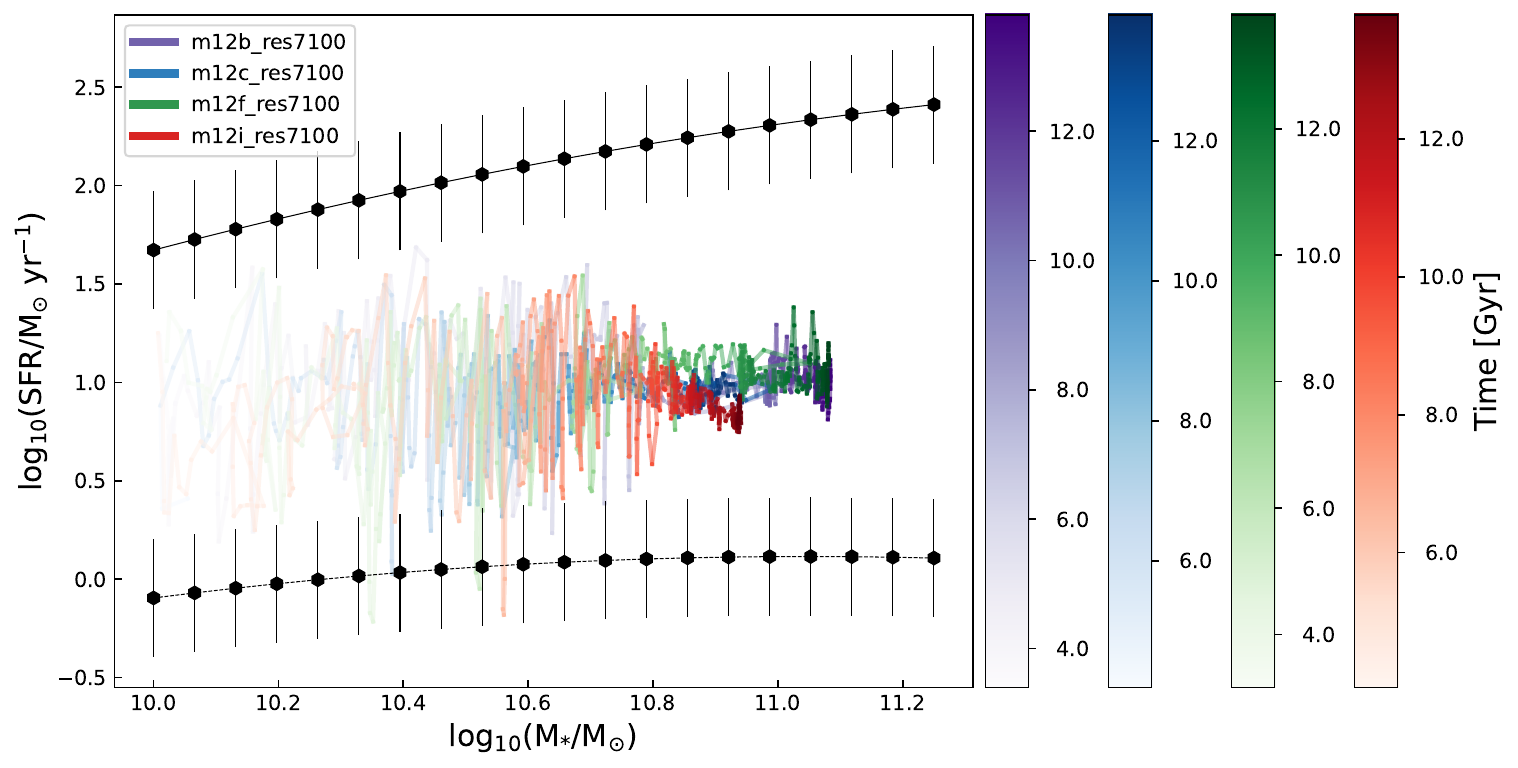}
\caption{Evolution of galaxies in log(SFR/M$_{\odot}$ yr$^{-1}$)-log(M$_{*}$/M$_{\odot}$) over time in the high-mass regime. Black lines correspond to the \cite{popesso2023} SFMS fit. Error bars show the accepted scatter of 0.3 dex.
\label{sfms_track_hm1}}
\end{figure}

We can visually see how the evolutionary behaviour of galaxies is dependent on their stellar mass content by looking at the SFMS tracks for each galaxy based on different stellar mass bins. In Figures~\ref{sfms_track_lm1},~\ref{sfms_track_im1},~\ref{sfms_track_hm1}, we show the evolution of a subset of galaxies in three mass bins. The mass bins are separated into low-mass (log(M$_{*}$/M$_{\odot}$) $<$ 9.0), intermediate-mass (9.0 $<$ log(M$_{*}$/M$_{\odot}$) $<$ 10.0), and high-mass (log(M$_{*}$/M$_{\odot}$) $>$ 10.0), which follow similar definitions to the mass bins in \cite{smith2024}.

\afterpage{\clearpage}

Similar to \cite{smith2024}, we find that galaxies in the lower-mass regime systematically lie below the SFMS throughout cosmic time. Higher-mass galaxies tend to lie above the SFMS. Most of the variability at intermediate masses is clustered towards SFRs below the SFMS, whereas low-mass variations typically lie above or near the main sequence, dropping below at later times. We do not observe an apparent flattening at the low-mass end, as proposed by some previous works \citep{barro2019}, but we again caution against over-interpreting a synthetic catalog as opposed to an observationally selected one.

Our fit at z = 0 is lower than the observed ones at intermediate and low masses, while at high redshift the fitted SFMS is lower than observed ones at all masses. Our differences in SFMS fits can be explained by the lack of high-mass, high-redshift systems, lack of massive quenched systems at low-redshift, and lack of early-type galaxy progenitors, which tend to have higher SFRs at earlier cosmic times \citep{sanchez2019}.

\subsubsection{Morphology}

Phases of spheroidal morphology are transient events that are only present for approximately 30 per cent of each galaxy's evolution, on average. Of the snapshots with successful S\'ersic fitting, 3060 snapshots were classified as disk-dominated and 1090 snapshots were classified as spheroidal-dominated. These phases tend to coincide with early cosmic times ($t_{universe} \lesssim 4$ Gyr) and lower stellar masses, typically accompanying a major merger event. This is a consequence of selecting z = 0 disks since the disks would not be present with extended periods of major merger activity.

\begin{figure}[hbt!]
\centering
\includegraphics[scale=0.33]{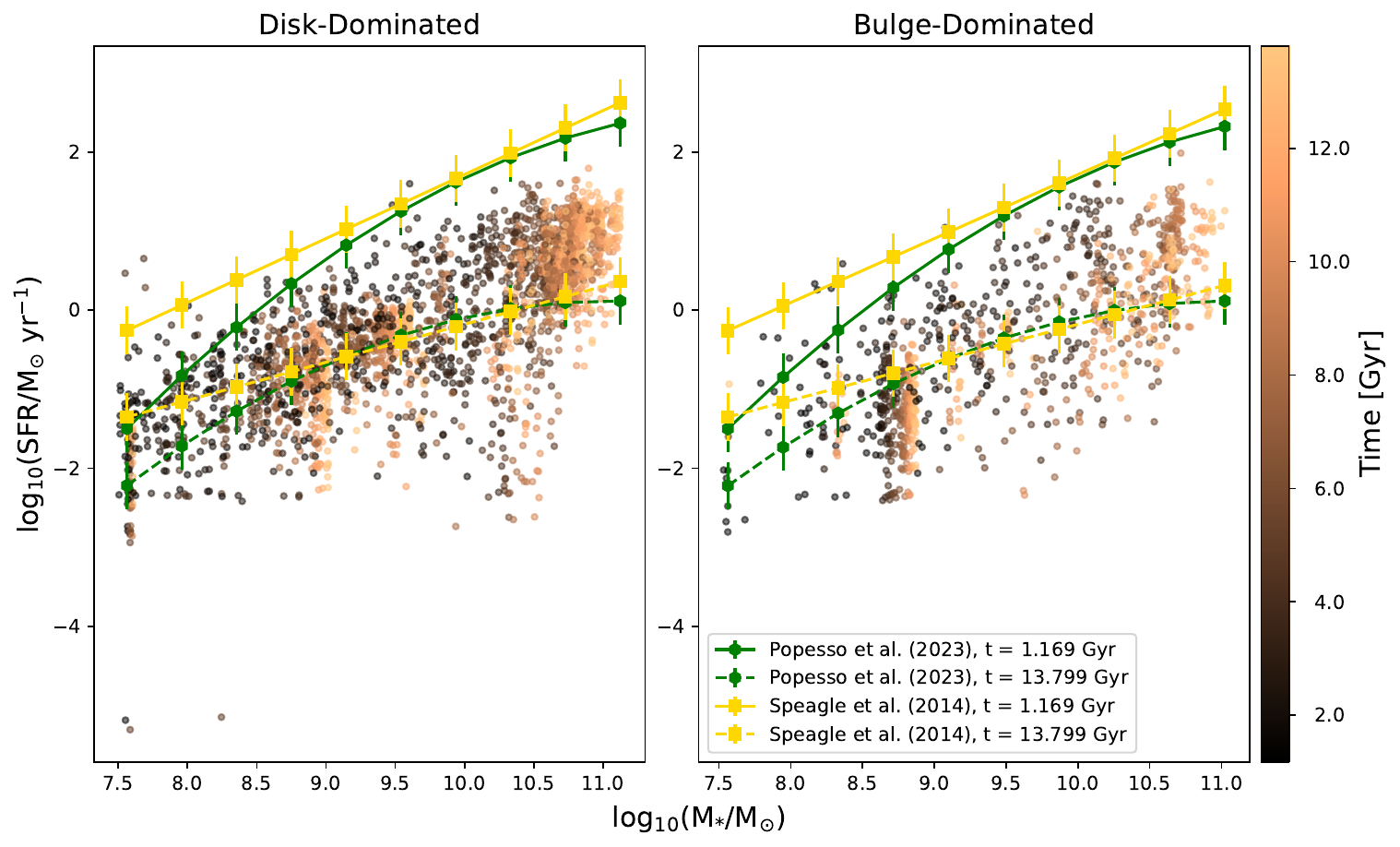}
\caption{SFMS relation for all galaxies analysed. On the left panel, snapshots where galaxies have n $<=$ 2.5 are shown whereas the right panel shows snapshots where n $>$ 2.5. Comparison to observed SFMSs are displayed with colored lines. Solid lines indicate the SFMS calculated at the earliest age of the Universe analysed and dashed lines show the expected SFMSs at z = 0.0. Error bars show the accepted scatter of 0.3 dex.
\label{sfms_split}}
\end{figure}

Figure~\ref{sfms_split} shows how the position of galaxies relative to the SFMS depends on the apparent S\'ersic index. The mass trends seen when considering the total sample are still present.

Galaxies typically spend more time in the disk-dominated regime, and so naturally exhibit a broader range of positions relative to the main sequence. Particularly, disks are more common at low redshift since sufficient mass infall and time result in most halos forming stable disks.

\subsection{SFMS Deviations}

\begin{figure}
\centering
\includegraphics[scale=0.33]{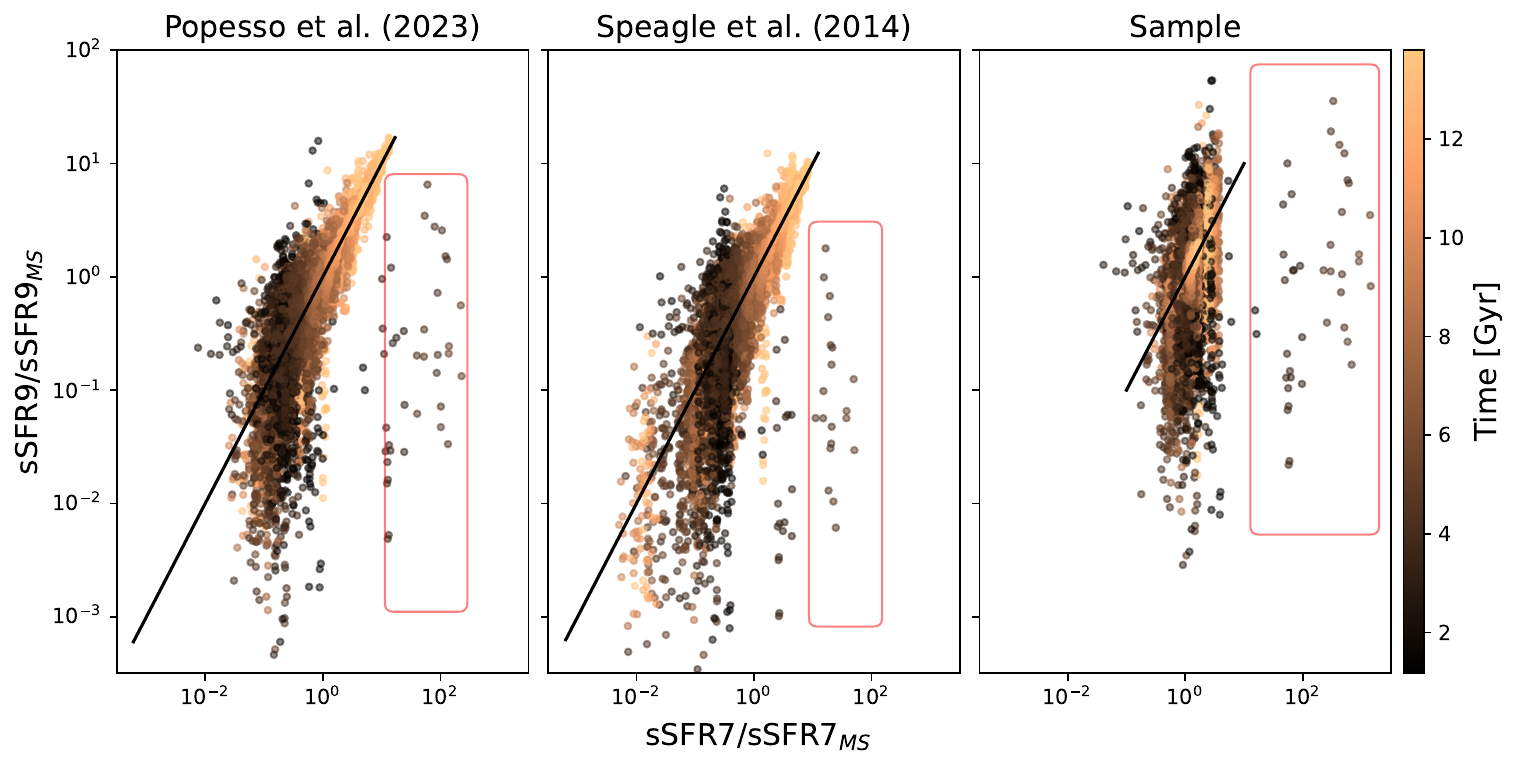}
\caption{Comparison between SFMS deviations on long (SFR9) and short timescales (SFR7). Black lines indicate equivalence. Red boxes surround outlier points from a Milky Way-Andromeda pair. SFMS deviations are plotted logarithmically for comparison with the SFMS.
\label{sfms_dev}}
\end{figure}

We next review deviations from the three SFMSs. Selection effects in the observed SFMSs are effectively ignored. In Fig~\ref{sfms_dev}, the SFMS deviations measured over different timescales compared to each expected SFMS is shown. Motivated by \cite{wang2020b}, we consider deviations of SFRs averaged over short timescales (on the order of 10 - 25 Myr depending on the time between snapshots, SFR7) and over long timescales (averaged over 800 Myr, SFR9). The deviations follow the forms below: 
\begin{equation}
\begin{split}
    \Delta sSFR7 = \ <sSFR_{gal}>_{10 \ \rm{Myr}} \\ 
-<sSFR_{SFMS}>_{10 \ \rm{Myr}} [yr^{-1}],
\end{split}
\end{equation}
\noindent
\begin{equation}
\begin{split}
\Delta sSFR9 = \ <sSFR_{gal}>_{800 \ \rm{Myr}} \\
- <sSFR_{SFMS}>_{800 \ \rm{Myr}} [yr^{-1}]. \\
\end{split}
\end{equation}
\noindent
While these quantities are evaluated in units of yr$^{-1}$, we consider deviations in log-space (as ratios) in Figure~\ref{sfms_dev} for easier comparison to the SFMS.

By investigating the correlations between short and long timescales, we can determine if any physical processes dominate these deviations. For example, if long-term deviations dominate then long-term processes such as accretion are important. Similarly, if short-term deviations dominate, then in-situ star formation processes such as supernovae-driven feedback are important. In addition, any separation in parameter space that may be present would imply a lack of ergodicity, since the initial position of a galaxy on the SFMS would determine its trajectory. 

The short-term deviations are more tightly packed within the range of -1.0 to 1.0 dex when compared to \cite{popesso2023} than when compared to \cite{speagle2014}. However, galaxies are more evenly spaced within the parameter space on both short and long timescales when compared to \cite{speagle2014}. There is an absence of points populating the high $\Delta$sSFR7 - $\Delta$sSFR9 parameter space, since high $\Delta$sSFR9 would require prolonged periods of enhanced SFR above the SFMS. 

For observed SFMS relations, there is an apparent trend with time: at early times, galaxies systematically lie below the SFMS resulting in negative deviations, and by later times galaxies systematically lie above the SFMS resulting in positive deviations. In fact, there is a strong, statistically significant correlation between the these deviations and time. However, these biases disappear when we consider our sample-constructed SFMS, as one would expect, since the SFMS considered is the zero-point reference for calculating $\Delta$sSFR. This trend is almost certainly caused by comparing a catalog of simulated galaxies to an observationally derived catalog. We note that the small cloud of points at earlier cosmic times and higher $\Delta$sSFR7 values are from one of the Milky Way-Andromeda halos, highlighted by red boxes. We also note the apparent increased scatter at later-times seen in sample deviations compared to the observed SFMS. That is, galaxies are more evenly distributed above and below the sample-constructed SFMS than observed SFMSs. The strong time correlation may also indicate a lack of diversity in our SFHs resulting from our sample construction. In addition, galaxies lying systematically above or below the SFMS presents possible tensions with observed systems that fluctuate about the SFMS regularly such as in \cite{camps2026}. However as noted by \cite{camps2026}, galaxies tend to remain on the same side of the SFMS for the majority of the past gigayear. We do see snapshots with close to zero deviation at intermediate masses and cosmic times, but we reiterate that the time correlation and resulting tension is due to comparing our simulated sample to observed relations. In summary, we use a SFMS fitted on our data due to our simulated sample not representing the observed SFMS fits.

\subsubsection{Quantifying the Bursty to Smooth SFR Transition}

\begin{figure}
\centering
\includegraphics[scale=0.58]{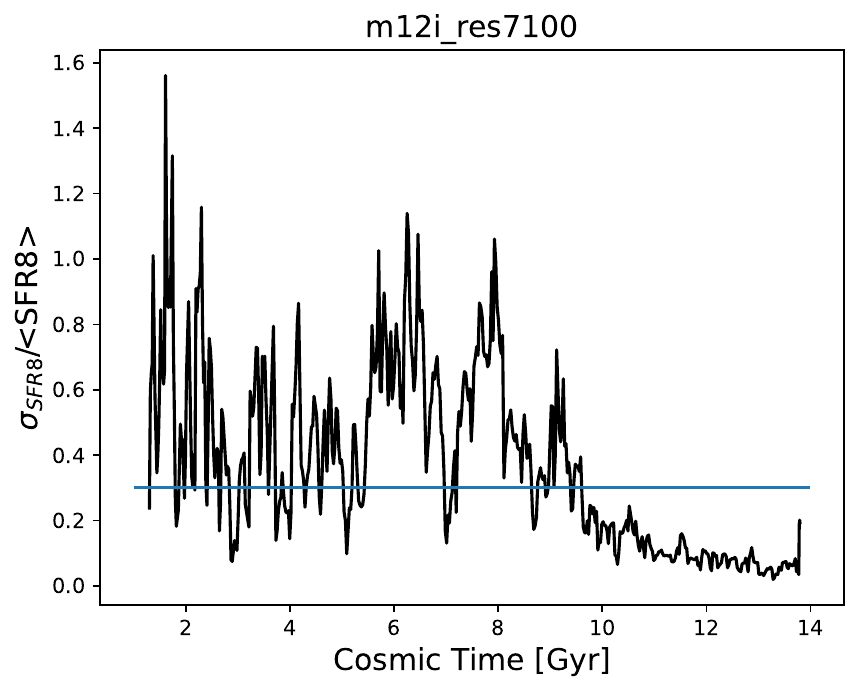}
\caption{Normalized moving standard deviation over m12i's SFH. The variance is calculated over the previous 100 Myr, for each analyzed snapshot. The standard deviation is normalized by the moving SFR average. The horizontal blue line denotes our adopted turnover threshold of 0.3 from bursty to smooth star formation that is motivated by the commonly accepted scatter of the observed SFMS. 
\label{turnover_fig}}
\end{figure}

In our higher-mass (Milky Way-like) systems, we see a profound and sudden transition from high SFR variability (i.e. bursty) to low SFR variability (i.e. smooth) at later times. This appears to follow the explanation of \cite{hop2023}, where the escape velocity of higher-mass systems becomes sufficiently large to prevent gas outflows from escaping the central galaxy. To quantify this transition, we measure the moving standard deviation over cosmic time in each of our 14 Milky Way-mass halos to constrain the times and halo masses that these apparent transitions occur. An example of this calculation is shown in Figure~\ref{turnover_fig}.

Considering only the halos that have normalized standard deviations below 0.3 (set to show what is within the accepted scatter of the SFMS) for extended periods (at least 1 Gyr), the average escape velocity of these transitions is about 200.82 km/s (log(M$_{halo}$ M$_{\odot}^{-1}$) $\sim$ 11.91), regardless of the length of the moving time window. Our results are broadly consistent (average halo mass difference of 29.75 per cent) with \cite{gurvich2023}, who analyzed 3 Milky Way-mass galaxies in the FIRE-2 sample using the archaeological approach. This estimate is also very similar to the escape velocity of 200 km/s estimated in \cite{hop2023}, however our estimate of the escape velocity is evaluated at the virial radius, not 10 - 20 kpc. This is not unexpected: to first order, we expect the radial profile of the escape velocity to flatten, similar to a rotation curve. While the total SFR variability is reduced after the observed transition, we do still see short-term variations occur on the order of 100 Myr or less, consistent with sub-gigayear variability \citep{camps2026}.

We also separate galaxies based on their spheroidal/disk morphologies to measure the impact of galaxy shape on short- and long-term SFMS deviations. On average, disk systems have a S\'ersic index of 1.23 and spheroidal morphology systems have a S\'ersic index of 6.28. When we consider the morphology of galaxies, we see that spheroid-dominated galaxies are restricted to a smaller range of short-term deviation values below the SFMS compared to disk-dominated galaxies. This is expected since spheroid-dominated galaxies are usually quenched. However, we note that this may also be due to how most galaxies spend the majority of their evolution in disk-dominated phases. Overall, we find that there is no apparent significant statistical difference.

\subsection{Measuring Ergodicity: TM Metric} \label{tm_metric_sect}

Ergodicity is quantitatively measured via the Thirumalai-Mountain (TM) metric \citep{thirumalai1989}. This metric is defined as the sum of the mean-squared deviation of each individual's time-average from the ensemble-averaged time-average,
\begin{equation}
\Omega_{e}(t) = \frac{1}{N} \sum^{N}_{i=1} [\epsilon_{i}(t) - \overline{\epsilon}(t)]^{2},
\end{equation}
\noindent
where $\epsilon(t)$ is the time average of a process over time $t$, $\overline{\epsilon}(t)$ is the sample average of all time averages, and $N$ is the size of the ensemble. When applied to SFMS deviations, the equation becomes

\begin{equation}
\begin{split}
    \Omega_{e}(t) = \frac{1}{N_{gal}} \sum^{N_{gal}}_{i=1} (<\Delta sSFR_{gal} - \Delta  sSFR_{SFMS}>(t) \\
- \overline{<\Delta sSFR_{gal} - \Delta sSFR_{SFMS}>}(t))^{2} \ \ [yr^{-2}],
\end{split}
\end{equation}
\noindent
where brackets denote time averages and the bar denotes averaging over the galaxy sample at each time $t$. For an ergodic system, we expect the TM metric to trend towards zero with steeper slopes associated with higher levels of ergodicity. We evaluate the TM metric in linear sSFR units of $ yr^{-2}$ and we use logarithmic axes to emphasize the shape of the TM metric behaviour over time in Sections~\ref{full_sample_tm} and~\ref{hist_depend}. In the following section, we comment on the limitations of the TM metric. 

\subsubsection{Limitations of the TM Metric}
\label{tm_limits}

To investigate the behaviour of the TM metric, we construct a heteroscedastic sample. For simplicity, we set the mean to 1.0 and randomly sample from a normal distribution (mean of 0.0, standard deviation of 1.0). In our constant variance model, we sample from this Normal Gaussian noise 100 times and add 1.0. For our decreasing variance model, we multiply the random noise by a decreasing power law function of $x^{-1}$ and then add this decreasing noise to our constant mean.

\begin{figure}
\centering
\includegraphics[scale=0.38]{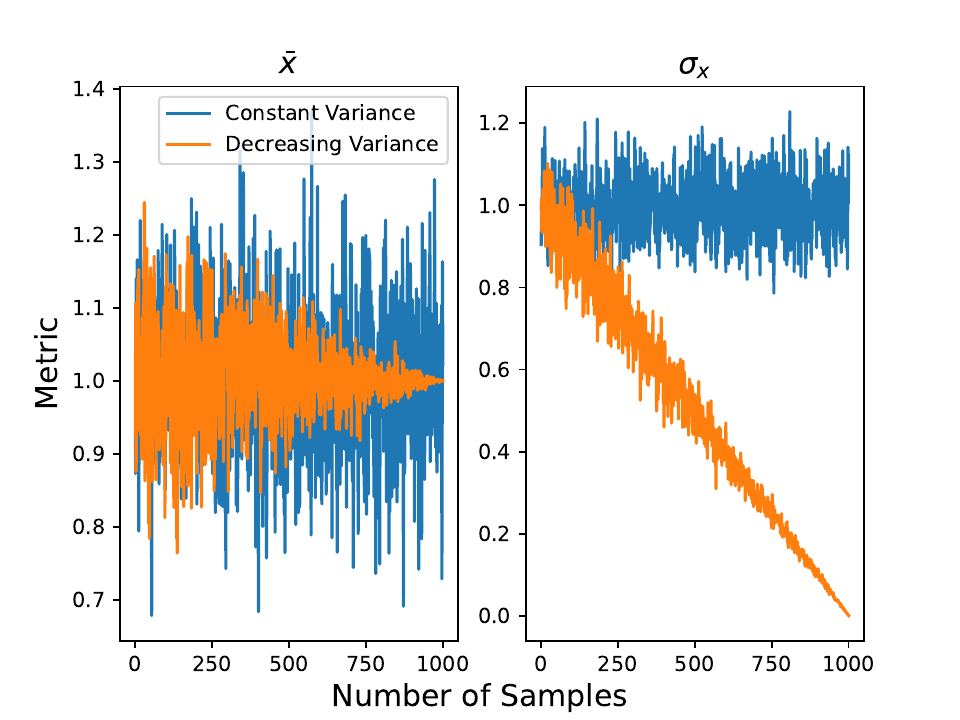}
\caption{Sample average and standard deviation of different noise models. The average of both samples are displayed in the left panel and the standard deviation of the samples is shown in the right panel. While both models produce averages around 1.0, only the normal noise model produces a mostly stationary variance. 
\label{appendix:app_2}}
\end{figure}

The shapes of the TM metric curves are nearly indistinguishable, with power law exponent fits of 0.91 and 1.02 for the constant variance and decreasing variance models, respectively. Despite the different variances, the shapes of the curves are strikingly similar. Interestingly, by the TM metric convergence alone, one might naively expect the constant variance process to be nonergodic and the decreasing variance process to be ergodic. However, when we consider the stationarity of both processes, shown in Figure~\ref{appendix:app_2}, it is obvious that the decreasing variance process cannot be ergodic. Hence, although a process may look ergodic by the TM metric, the process can still fail to be ergodic.

Despite the different variance behaviours the variance of the running time-averages are almost indistinguishable, particularly for large samples.

\subsubsection{TM Metric for the Full Sample}
\label{full_sample_tm}

\begin{sidewaysfigure*}[hbp!]
\centering
\vspace*{-9cm}
\hspace*{0cm}\includegraphics[scale=0.6]{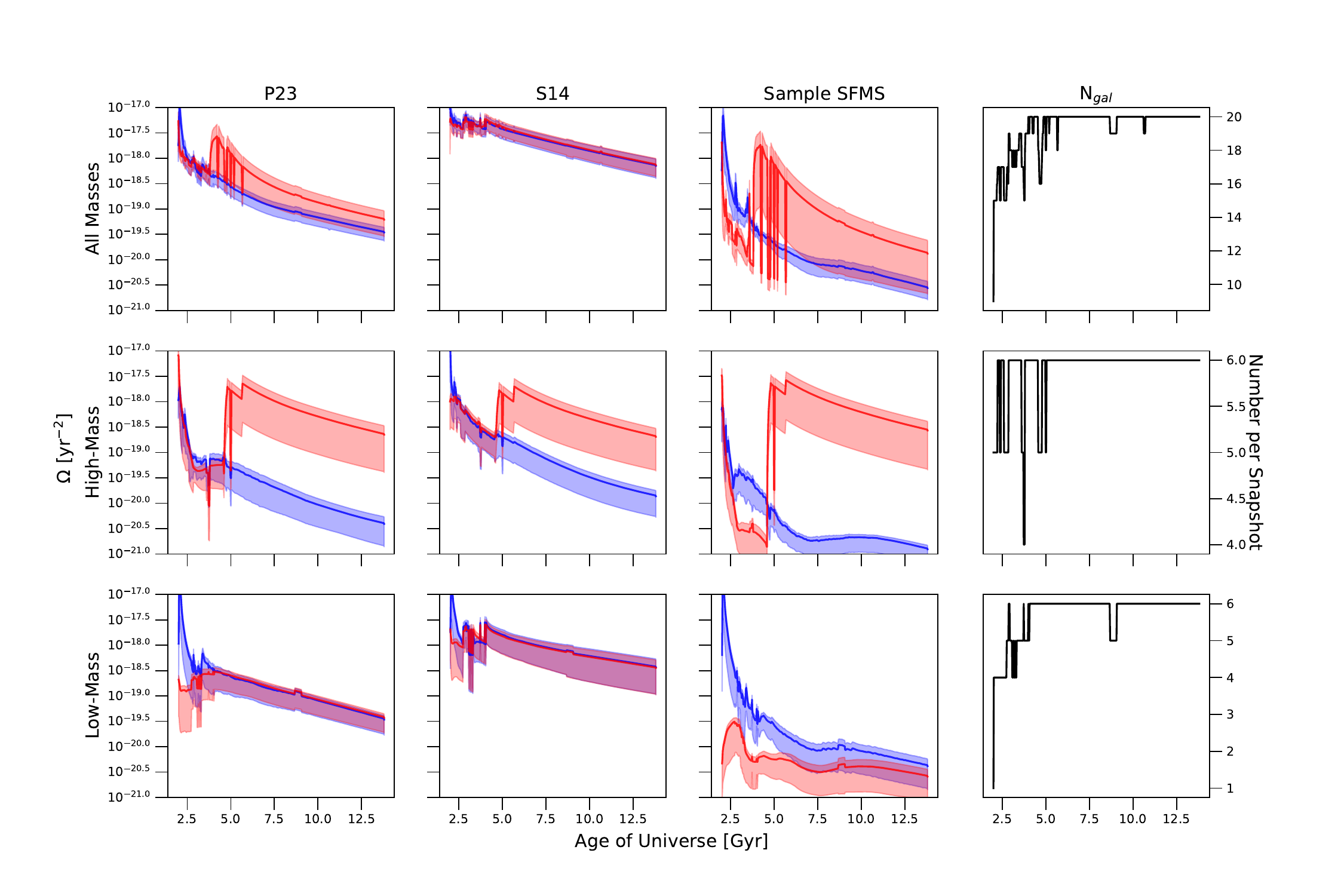}\hspace*{0cm}

\caption{TM metric over cosmic time across our sample for SFMS deviations. In the first three columns, the TM metric based on SFMS deviations over cosmic time is shown for the three SFMSs considered in this work. Red lines denote SFR deviations on long timescales (800 Myr, sSFR9) and blue lines are for short timescales ($\sim$20-25 Myr, sSFR7). The rightmost column contains the number of galaxies considered at each snapshot. Galaxies may not be considered during a snapshot due to lack of star formation and/or the halo was not captured by the analysis code. The top row of panels corresponds to the whole sample, and the middle and bottom rows show the TM metric behaviour when only the 6 highest- and lowest-mass systems are considered, respectively. Logarithmic axes for the TM metric are used to improve clarity.
\label{tm_metric_sfms}}
\end{sidewaysfigure*}

We calculate the TM metric with respect to the vertical SFMS deviations from each observed SFMS. These results are shown in Fig~\ref{tm_metric_sfms}. The analysis code may miss a galaxy at any given snapshot due to a lack of star formation or failure to converge on halo properties. As a result, the number of galaxies analysed at each snapshot does not always increase.

The TM metric for SFMS deviations trends towards zero when either of the observed SFMS is considered, as well as the sample SFMS. Regardless of the main sequence used, the TM metric of SFMS deviations all appear to trend to zero with cosmic time (we examine convergence rates in Section~\ref{tm_conv_sec}), implying an approach to ergodicity. In addition, this is seen both in terms of short-term deviations and long-term deviations. The initial rises and falls in the $\Delta$sSFR9 data is caused by high-variance galaxies being added to the sample which alter the ensemble average and increase the variance in addition to the count variation at early times. This is more visible with the long-term deviations due to the nature of moving-averages with larger averaging-windows being more sensitive to changes in the variance. The $\Delta$sSFR7 TM metric may drop at early times due to changes in the sample size and/or feedback temporarily shutting off star formation in lower-mass systems. We do note that the TM metric is inherently larger (on average 1.17 dex for short-term deviations, 0.47 dex for long-term deviations) when considering the \cite{speagle2014} SFMS, which is due to the bias of the fit to our data at low-masses and high-masses. This bias more heavily impacts the short-term deviations, resulting in the TM metric curves for $\Delta$sSFR7 and $\Delta$sSFR9 to have similar normalizations. The lack of a high-mass turnover is a known issue with the \cite{speagle2014} fit and further motivates our comparison with the SFMS fit in \cite{popesso2023}.

Given the burstiness of SFHs in FIRE-2, these apparent tendencies toward ergodicity are not surprising: galaxies could be expected to explore more of the log(SFR)-log(M$_{*}$) parameter space over time versus a system with smoother changes in the SFR. Furthermore, higher-mass systems converge to similar positions ($\sigma \approx 0.071$ dex in average $\Delta$sSFR7 values for log(M$_{*}$ M$_{\odot}^{-1}$) $\geq$ 10.0) relative to the SFMS at later times, resembling stationary behaviour. While stationarity in the ensemble mean is essential for ergodic systems, individuals exhibiting stationarity in SFMS deviations would actually result in a lack of ergodicity due to insufficient individual variation and convergence to static SFRs. In terms of similarity of outcomes, when considering the simulated data-constructed SFMS, the TM metric behaves more similarly to the \cite{popesso2023} SFMS, except with larger values for $\Delta$sSFR9 and smaller values for $\Delta$sSFR7. Nevertheless, once beyond the early growth period, we find that the general slopes over time are robust regardless of the SFMS fit used for our sample and the star formation indicator used.

\subsubsection{TM Metric Separated by Mass}

The core sample of FIRE-2 simulations contains predominately Milky Way-mass galaxies at $z = 0$. In order to constrain any mass-dependence on our results, we consider the ergodicity of SFMS deviations in our 6 lowest-mass and 6 highest-mass systems over cosmic time. The TM metric behaviour for these two groups are shown in Figure~\ref{tm_metric_sfms}. Overall, the TM metric behaves similarly to the total sample over time (with the convergence rates of these mass-separated groups presented in Section~\ref{tm_conv_sec}).

\subsubsection{TM Metric Convergence}
\label{tm_conv_sec}

As a further constraint on measuring ergodicity, we consider the convergence of the TM metric towards ergodicity, following the relation proposed in \cite{thirumalai1989} and fitting the convergence in a manner that has been applied to various systems exhibiting both ergodic and non-ergodic behaviours \cite[e.g.][]{suzen2014}. From \cite{thirumalai1989}, it is proposed that for diffusive systems,
\begin{equation}
    \Omega_{e}' (t)= \frac{\Omega_{e}(t)}{\Omega_{e}(0)} \sim \frac{t^{-\alpha}}{D_{e}}
\end{equation}

\noindent
where $D_{e}$ is related to the time required to reach ergodicity and $\alpha=1$ for diffusive systems. Thus, we can then compare the rate of ergodic convergence to the expected convergence in the diffusive regime. Specifically, we can measure the rate of ergodic convergence by fitting TM metric convergence with a power law. Power law exponents of unity ($\alpha = 1$) imply behaviour similar to diffusive systems (i.e. ergodic), exponents above unity ($\alpha > 1$) imply the system is super-diffusive (akin to Brownian motion and is considered ergodic), and exponents below unity ($\alpha < 1$) indicate the system is sub-diffusive (does not attain ergodicity). A discussion of these behaviours and analysis is provided in \cite{baccetti2024}. The power law exponents for each SFMS fit and each mass group are given in Table~\ref{plaw_tab}.

\begin{table*}
\centering
\resizebox{\linewidth}{!}{
\begin{tabular}{|c | c c | c c | c c|}

\hline
Mass Group & \multicolumn{2}{c|}{P23} & \multicolumn{2}{c|}{S14} & \multicolumn{2}{c|}{Sample} \\

& $\alpha_{sSFR7}$ & $\alpha_{sSFR9}$ & $\alpha_{sSFR7}$ & $\alpha_{sSFR9}$ & $\alpha_{sSFR7}$ & $\alpha_{sSFR9}$ \\

\hline
 Complete (7.5 $\leq$ log(M$_{*}$/M$_{\odot}$) $<$ 11.5) & 4.78 $\pm$ 2.38 & 1.31 $\pm$ 5.30 & 0.97 $\pm$ 0.25 & 0.54 $\pm$ 0.19 & 8.36 $\pm$ 2.53 & 5.25 $\pm$ 17.14 \\
Low (log(M$_{*,z=0}$/M$_{\odot}$) $<$ 9.8) & 7.19 $\pm$ 4.82 & 0.61 $\pm$ 1.89 & 3.47 $\pm$ 3.14 & 0.70 $\pm$ 0.51 & 7.28 $\pm$ 4.37 & 1.07 $\pm$ 0.78 \\
High (log(M$_{*,z=0}$/M$_{\odot}$) $>$ 10.9) & 8.89 $\pm$ 5.61 & 8.38 $\pm$ 1.44 & 21.78 $\pm$ 14.49 & 0.86 $\pm$ 0.97 & 13.80 $\pm$ 8.57 & 10.31 $\pm$ 18.08 \\

Complete (500 Myr Blocks) & 0.31 $\pm$ 0.07 & 0.55 $\pm$ 0.14 & 0.23 $\pm$ 0.13 & 0.21 $\pm$ 0.11 & 0.62 $\pm$ 0.11 & 0.62 $\pm$ 0.18 \\
Complete (1 Gyr Blocks) & 0.64 $\pm$ 0.14 & 0.21 $\pm$ 0.21 & 0.37 $\pm$ 0.15 & 0.34 $\pm$ 0.13 & 1.13 $\pm$ 0.09 & 0.10 $\pm$ 0.31 \\

\hline
\end{tabular}
}
\caption{Power law exponents derived from TM metric convergence fit using the 3 SFMSs analyzed. Slopes above unity are indicative of ergodic convergence, slopes close to unity imply Brownian motion, and slopes below unity imply non-ergodic behaviour. The two bottom rows are exponents fitted onto the complete data set after block scrambling with block sizes given in brackets.}
\label{plaw_tab}
\end{table*}

There is apparent ergodic convergence in vertical SFR7 deviations seen in almost all mass groups and SFMS fits considered. This makes intuitive sense since short timescale and larger amplitude variations about the mean that dampen at later times will result in individual galaxies exploring the full deviation parameter space before converging to values similar to the ensemble average. One exception is the near linear fit for the full sample using the SFMS from \cite{speagle2014}. On longer timescales, the SFR9 SFMS deviations only appear to lie in the super-diffusive regime at stellar masses $\rm{log}(M_{*}/M_{\odot}) \geq 9.0$, and only when using the SFMS constructed from the FIRE-2 data. However, this behaviour may be attributed to the \cite{speagle2014} fit lacking a high-mass turn-off. Hence, the ergodic convergence of longer timescale deviations may be more sensitive to stellar mass and the SFMS used. We note that this may also be due to the smaller data set of snapshots with lower-masses relative to the overall evolution.

\subsection{Testing History Dependence via Block Scrambling}
\label{hist_depend}

\begin{figure}
\centering

\includegraphics[scale=0.26]{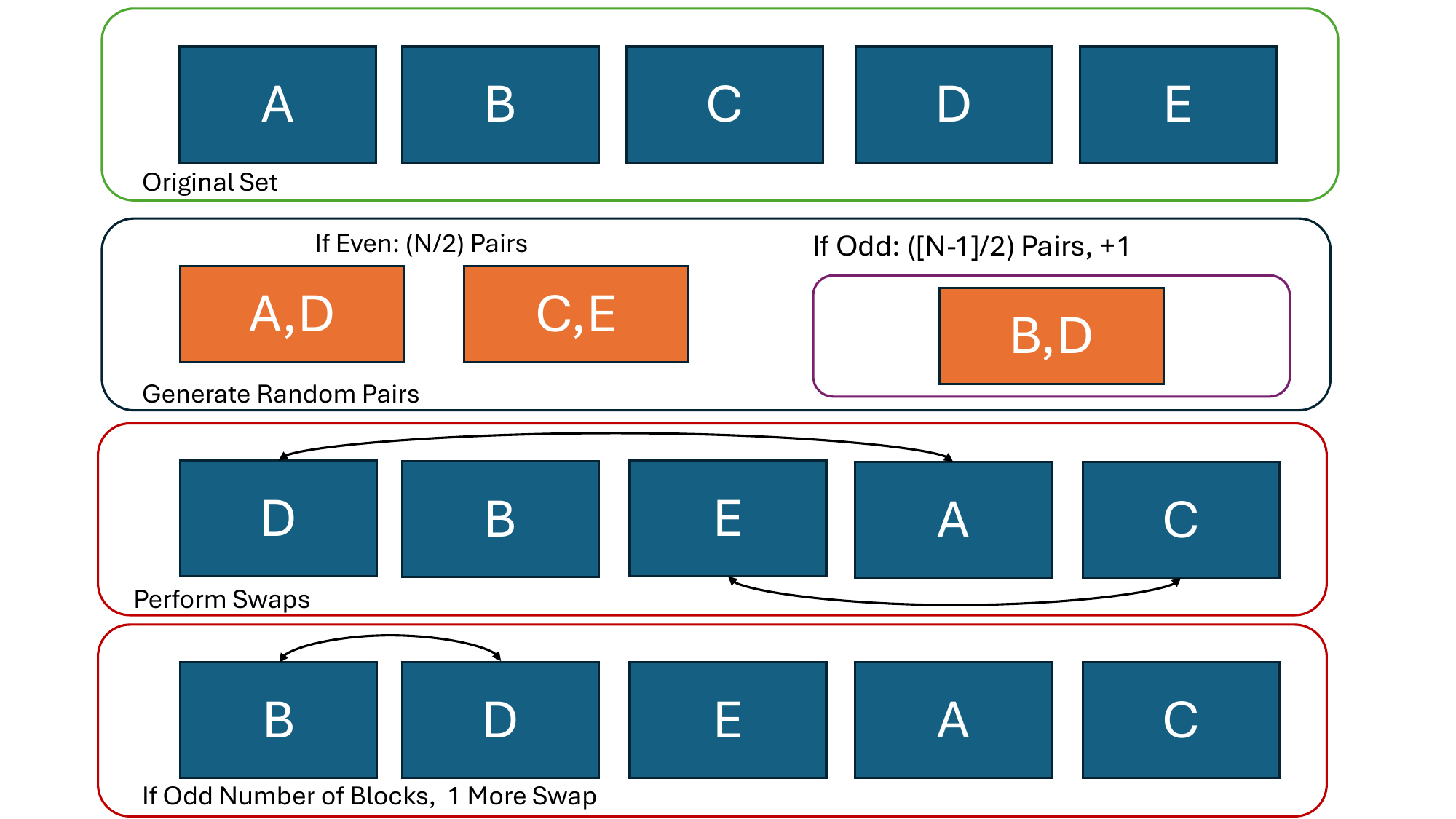}
\caption{Diagram visualizing the pairwise block-scrambling routine used in this work. For an odd number of blocks, an extra swap is performed to ensure all blocks occupy positions different from their original.
\label{block_scramble_diagram}}
\end{figure}

\begin{figure}
\centering

\includegraphics[scale=0.38]{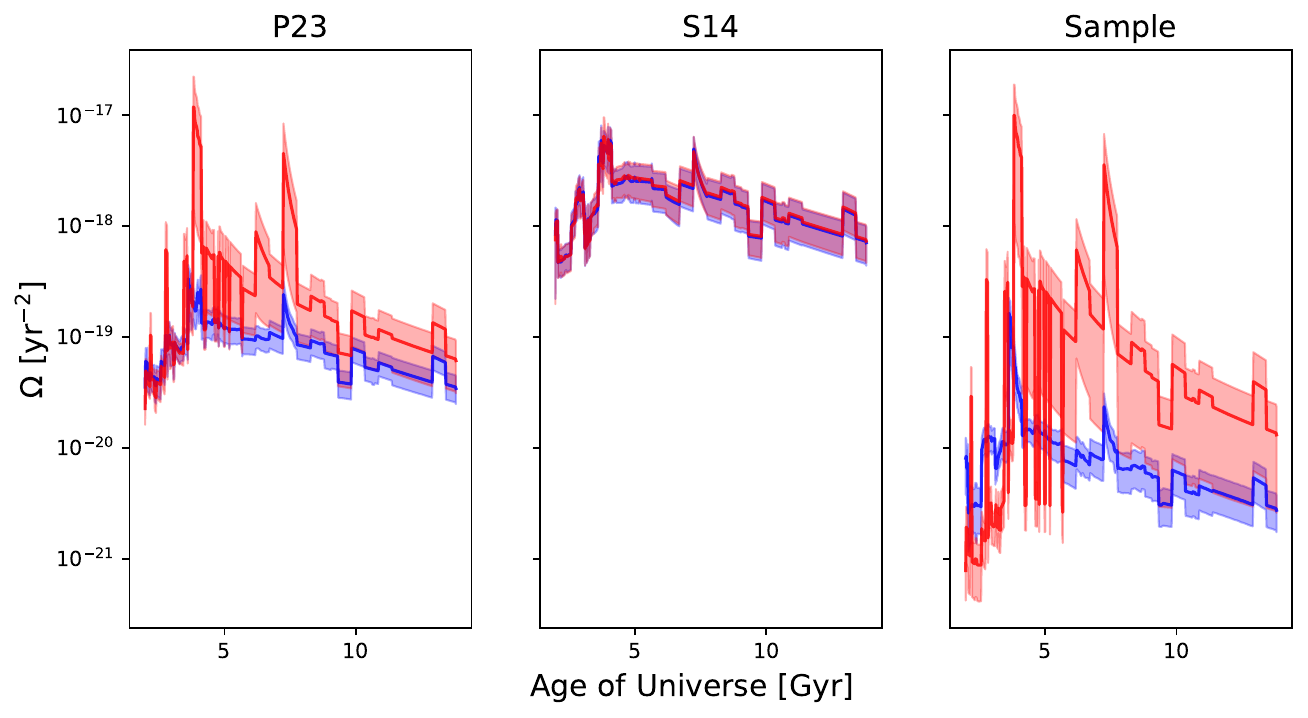}
\caption{TM metric as a function of cosmic time after block-scrambling. Columns separate different SFMS forms considered. The red lines denote SFR deviations on long timescales (800 Myr, sSFR9) and the blue lines are for short timescales ($\sim$20-25 Myr, sSFR7). The time series is divided into 500 Myr blocks and the maximum number of unique, non-overlapping block-scrambles is performed. Logarithmic axes for the TM metric are used to improve clarity.
\label{tm_block_scramble_eg}}
\end{figure}

To investigate any potential systematic temporal trends in the data, we separated the analyzed time series into equal-sized, non-overlapping blocks and shuffled the order of the blocks and repeated calculations of the TM metric on the scrambled data. To create a block scrambled version of the time series we randomly select a pair of blocks and then swap, repeating the process until all blocks have been swapped with another (random) block. A diagram visualizing this process is shown in Figure~\ref{block_scramble_diagram}. As discussed in \cite{kelty2022}, this removes the temporal information and can be used to determine the effects of temporal correlation on ergodicity. It can also be reasonably objected that the length of blocks chosen in the scramble may influence outcomes as well due to edge effects. To this end we have analyzed two block sizes: shorter and longer (500 Myr and 1 Gyr, respectively). We have limited sampling in block lengths less than 500 Myr to keep some portrayal of the normal physical evolution of SFRs. An example of the TM metric as a function of cosmic time after block-scrambling is shown in Figure~\ref{tm_block_scramble_eg}. We note that there are unavoidable visible edge effects in the block-scrambled data series, however the overall behaviour of the block-scrambled data is unaffected. For our calculated SFHs, across all SFMSs considered, the TM metric does not show behaviour associated with ergodic systems once time correlations are removed, regardless of the time block size. When considering archaeological SFHs we find similar results.

There is a clear difference in TM metric behaviour when the time series are block-scrambled. Interestingly, we find that the TM metric asymptotically approaches a nonzero value when the time series of SFMS deviations is block scrambled, evidenced by convergence values all below 1.0. The convergence values for both time block sizes using each SFMS are shown in Table~\ref{plaw_tab} in the bottom two rows. We also note the reduction in the separation between $\Delta$sSFR7 and $\Delta$sSFR9 curves compared to the regular time-series data. By removing the history-related trends present in the data, the overall variance of the ensemble is reduced and thus produces lower TM metric values. The TM metric calculated with the \cite{speagle2014} SFMS appears to be less impacted by temporal biases since the TM metric values are similar before and after block-scrambling.

\section{Discussion}

These results demonstrate the subtle behaviours and history-related issues that are inherent in attempting to classify ergodicity via a metric approach. We have shown that while on first appearances SFMS deviations appear to behave ergodically when plotted in their natural time sequence, in fact these deviations are not truly ergodic when considered under time-based scrambles. In ergodic systems block-scrambling should preserve the ergodic behaviour as is seen in other works on ergodic systems. For example, the TM metric shows asymptotic behaviour for time-correlated data unless the time series is block scrambled in \cite{kelty2022} and \cite{mangalam2023}. This can be equally taken as a weakness in the TM-metric: that is, the TM metric decreases with time due to the SFMS deviations having a time-dependence that mimics ergodicity in the TM metric rather than intrinsic ergodic behaviour (i.e. apparent but not true ergodicity). The origin of this behaviour is ultimately that the sample deviations do not show strong-sense stationarity. The convergence of the TM metric to zero requires the ensemble average to be constant, but not the sample variance (see Section~\ref{tm_limits}). Thus, block-scrambling data and recalculating the TM metric is an important and straightforward test to confirm or deny apparent ergodic behaviour.

The results found do agree with some observational studies. For instance, the strong time-dependence seen in this work has been noted by other studies where galaxies tend to remain in similar positions relative to the SFMS on Gyr timescales \citep{arango2025,wan2025,camps2026} or that present-day galaxies above the SFMS tend to have lower SFRs at early times, and vice versa \citep[e.g.][]{citro2016,sanchez2019,scholz2023}. However, there is also evidence for galaxies regularly crossing the SFMS \citep[e.g.][]{tacchella2018,camps2026}, which is not seen in our most massive systems. This potential tension can be explained by the nature of our simulated sample: the systems we investigate are specifically massive halos with prominent disks at z = 0, and hence do not experience major quenching events. Thus, it is important to re-emphasize that our results apply specifically to currently star-forming galaxies. As noted by other authors \citep[e.g.][]{sanchez2019,paola2022,stephenson2024}, morphology and mass are important factors for the quenching of star formation and would produce a larger variance in our SFHs, possibly diminishing the observed time bias. Interestingly our results may apply to progenitors of quiescent systems, since evidence for such progenitors to lie on the SFMS has been seen previously using various methods \citep{genel2018,moster2020,bosi2025}. Extrapolation to quenched systems requires further research.

Apparent ergodic convergence based on the TM metric has been previously observed. The general behaviour of the TM metric decreasing with time can be seen in scaled Brownian motion, which is heteroscedastic, non-stationary, and has been shown to exhibit ergodicity breaking \citep{safdari2015}. In other words, a non-ergodic process appears to exhibit ergodicity (i.e. apparent ergodicity) using the TM metric. \cite{cherstvy2015} have shown that heterogeneous diffusion processes, while non-ergodic over long time periods, the TM metric can show behaviour similar to Brownian motion in shorter time limits. Modified metrics for measuring ergodicity in complex systems have been proposed \citep[e.g.][]{lanoiselee2016,kelty_stephen2023} and can be utilized as a further ergodicity test after initial tests using the TM metric. These metrics will be investigated in future work.

These results actually build on our previous work in \cite{smith2024} in that we now have further understanding of how while the TM metric is a useful test for ergodicity, there are notable issues to be considered in its application. In \cite{smith2024} we found that constant SFRs produce stationarity but are insufficient for the application of the TM metric. Here, we show that even with variations there can be history-related issues (specifically a reduction in the variance) that can mimic ergodicity while not truly being ergodic. We also note that the issue of SFMS deviations having a temporal bias may be weakened and thus approach more ergodic behaviour at certain restricted epochs (e.g. cosmic noon) in contrast to the long evolutionary epochs considered in this work. If we restrict our sample to only consider high-redshift galaxies, we inevitably remove the observed transition from bursty to smooth SFHs and hence remove the associated change in $\Delta$SFR observed here. This work has been limited to redshifts $z \leq  4$ to maximize the available number of snapshots in our sample. Investigation into high-redshift systems is beyond the scope of this work but is a next step in the analysis.

\section{Conclusions}
\label{Conclusions}
In this study, we investigate the ergodicity of SFMS deviations for galaxies in the FIRE-2 project. With explicitly resolved feedback, we are able to better constrain how stellar feedback and major mergers affect ergodicity of SFMS deviations. Below we summarize our main findings:

\begin{itemize}
    \item The observed variability of SFHs gives credence to the notion that the SFMS is a result of short-term stochasticity, as opposed to a simple population average. Bursty star formation naturally explains observed scatter at lower-masses/high-redshifts, consistent with previous work \citep{cole2025}. At higher-masses and later times, the steady, smooth SFHs align with how we naively expect galaxies to evolve along the SFMS.

    \item Spheroid-dominated galaxies have smaller ranges of SFMS deviations than disc-dominated systems. Note phases of predominately bulge morphology in this study are transient events, mostly occurring at early cosmic times. 
    
    \item There is a clear trend with time and the values of SFMS deviations, both in short-term and long-term deviations. We see that at early times/lower-masses, galaxies tend to lie below the SFMS (negative SFMS deviations) and at late times/higher-masses, galaxies tend to lie above the SFMS (positive SFMS deviations). This is likely due to the construction of our sample.

    \item Based on the TM metric and its convergence, FIRE-2 galaxies appear to exhibit ergodic behaviour in short-term deviations from the SFMS over cosmic time. This apparent ergodicity is seen in both low-mass and high-mass systems.  Longer timescale deviations from the SFMS may appear to follow similar TM metric behaviour at later cosmic times, however these systems do not appear to converge to an apparent state of ergodicity over cosmic time. The burstiness of FIRE-2 SFHs is crucial to exhibit ergodic properties in galaxy samples. This reiterates the importance of variability in SFR in studies of galaxy evolution compared to the historical picture of relatively smooth SFHs.  Consistent with \cite{smith2024}, without sufficient variability galaxies cannot be ergodic.

    \item While exhibiting apparent ergodicity, we find that the observed TM metric behaviour and resulting convergence is due to decreasing variance with time. In particular, the decrease in variance with time resulting from the transition from bursty to smooth star formation and lack of late mergers prevents SFMS deviations in this sample from being truly ergodic. By block-scrambling the time series of the data and removing temporal bias, ergodic convergence is not attained. Thus, we caution that while TM metric convergence to zero is necessary for an ergodic system, it is not a sufficient condition.
    
\end{itemize}

Future work will include AGN feedback to constrain how ergodicity is influenced by the presence of supermassive black holes. Although star formation is more suppressed in the presence of AGN, the sufficient variability of SFHs at early times is expected to preserve the apparent ergodicity seen here. With more restricted SFR values at late times in massive systems due to negative AGN feedback, the ensemble average may converge more quickly than in the absence of AGN. That is, AGN feedback can introduce massive quenched systems into our analysis. Negative AGN feedback is expected to decrease the overall variance of SFMS deviations in the most massive systems, however positive AGN feedback may still have local effects \citep{mercedes2023}. In cases where the overall star formation of massive systems is suppressed due to AGN feedback, non-ergodic outcomes similar to \cite{smith2024} may become common due to insufficient individual variation. Thus, AGN in massive systems could introduce breaking of ergodic behaviour. In addition, at high redshift, we expect positive AGN feedback to have a more profound effect than negative AGN feedback where positive AGN feedback can trigger bursts of star formation \citep{silk2024}.

\begin{acknowledgments}
We thank the anonymous Reviewer for their insightful comments that improved the quality of this paper. We thank the FIRE collaboration, especially Claude-André Faucher-Giguère, Chris Hayward, Phil Hopkins, Dušan Kereš, Eliot Quataert, and Andrew Wetzel, for making the data publicly available. We would also like to thank Ivana Damjanov, Marcin Sawicki, Rosa Mérida, and the SMU Extragalactic research group for insightful comments. Funding from National Sciences and Engineering Research Council of Canada, Canada Research Chairs Programme, and Research Nova Scotia are acknowledged. FMS acknowledges support from the Saint Mary’s Faculty of Graduate Studies and Research Graduate Awards, the Father Burke Gaffney Memorial Scholarship, and the John Despard de Blois Scholarship for his PhD degree. 

The analysis presented in this work made use of the \textsc{python} packages \textsc{NumPy} \citep{harris2020}, \textsc{SciPy} \citep{virtanen2020}, and \textsc{Matplotlib} \citep{hunter2007}.
\end{acknowledgments}

\section*{Data Availability}
A public version of the \textsc{gizmo} code is available at \url{http://www.tapir.caltech.edu/~phopkins/Site/GIZMO.html}. FIRE data releases are publicly available at \url{http://flathub.flatironinstitute.org/fire}. Codes for calculating the TM metric and performing block-scrambling are available from the author upon reasonable request.

%

\vspace{5mm}


\software{GIZMO \citep{hopkins2015},  
          GizmoAnalysis \citep{wetzel2016,wetzel2020},
          HaloAnalysis \citep{wetzel2016,haloanalysis}
          }

\bibliography{ref}{}
\bibliographystyle{aasjournal}



\end{document}